\documentclass{PoS}

\title{Prototype Front-end ASIC for Silicon-strip Detectors of J-PARC Muon g-2/EDM Experiment}

\ShortTitle{Front-end ASIC for J-PARC Muon g-2/EDM Experiment}

\author{\speaker{Yuki Tsutsumi}$^a$, Tetsuichi Kishishita$^b$, Yutaro Sato$^b$, Masayoshi Shoji$^b$, Manobu M. Tanaka$^b$, Tsutomu Mibe$^b$, and Junji Tojo$^a$\\
        a. Kyushu University, Department of Physics, 744 Motooka, Nishi-ku, Fukuoka, 819-0395, Japan\\
        b. KEK High Energy Accelerator Research Organization, 305-0801, 1-1 Oho Tsukuba Ibaraki, Japan
        E-mail: \email{tsutsumi@epp.phys.kyushu-u.ac.jp}}

\abstract{
We report on the development of a front-end ASIC for silicon-strip detectors of the J-PARC Muon g-2/EDM experiment. This experiment aims to measure the muon anomalous magnetic moment and electric dipole moment precisely to explore new physics beyond the Standard Model. 
Since the time and momentum of positrons from muon decay are key information in the experiment, a fast response with high granularity is demanded to silicon-strip detectors as the positron tracker.
The readout ASIC is thus required to tolerate a high hit rate of 1.4 MHz per strip and to have deep memory for the period of 40 us with 5 ns time resolution. To satisfy the experimental requirements, an analog prototype ASIC was newly designed with the Silterra 180 nm CMOS technology. 
In the evaluation test, the time-walk was demonstrated to reach 0.8~ns with a sufficient dynamic range of 6~MIPs and pulse width of 45~ns for 1 MIP event. 
The design details and performance of the ASIC are discussed in this article.
}

\FullConference{Topical Workshop on Electronics for Particle Physics (TWEPP2018)\\
		17-21 September 2018\\
		Antwerp, Belgium}

\begin{document}
\maketitle
\flushbottom

\section{Introduction}
The muon anomalous magnetic moment $(g-2)_{\mu}$ and the electric dipole moment are sensitive to new physics beyond the Standard Model (SM) \cite{miller}. In the meanwhile, there is a discrepancy between the SM prediction and the measurement by the E821 collaboration at Brookhaven National Laboratory (BNL) at more than $3~\sigma$ \cite{benett, keshavarzi, davier, jegerlehner, benayoun}. To make a definitive conclusion, a new measurement of the $(g-2)_{\mu}$ and the electric dipole moment (EDM) is under preparation based on a innovative technique using a muon beam at the Japan Proton Accelerator Research Complex (J-PARC). The experimental goals are to measure $(g-2)_{\mu}$ with a precision of 0.1 parts per million (ppm), which corresponds to the improvement from the BNL experiment by factor 5, and at the same time, to seek for the EDM with a sensitivity of $10^{-21}e\cdot$cm. The ultra-cold muon beam will be used at J-PARC, which is produced from high intensity proton beam by thermal muonium production, followed by laser ionization and muon linear acceleration.

\begin{figure*}
\centering
\includegraphics[width=6in]{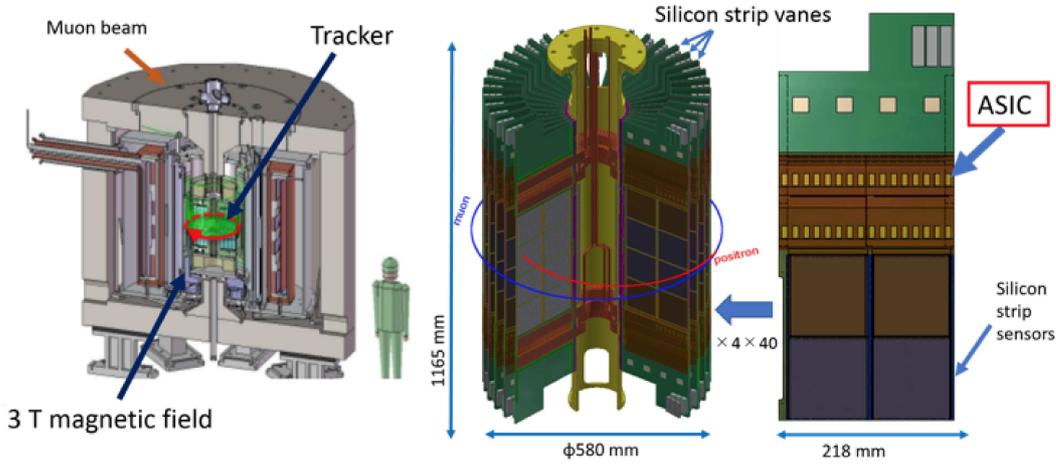}
\caption{Detector concept of the muon g-2/EDM experiment at J-PARC.}
\label{fig_g2}
\end{figure*}

The detector concept and geometry are shown in Figure \ref{fig_g2}. The muons are stored under a 3 T magnetic field following the orbital cyclotron motion. A silicon tracker is placed in a 66~cm-diameter compact muon storage ring for the detection of positrons from muon decay. 
 The 40 silicon strip vanes are currently planned to be aligned radially in the detection volume. Each vane has single-sided p-on-n type strip sensors on both sides with mutually orthogonal strips. Two-dimensional position of a positron track is therefore detected by two layers of the strip sensor.
To detect circular positron tracks from the $\mu^{+}\rightarrow e^+\nu_e \bar{\nu}_{\mu}$ decay, a fast response with high granularity is required to the silicon strip detectors in the storage magnet. Front-end ASICs have been developed for this purpose.

\section{Design of the prototype ASIC ``SliT2017TEG''}
\begin{figure*}[!t]
  \centering
    \begin{tabular}{c}

      \begin{minipage}{0.3\hsize}
          \includegraphics[keepaspectratio, scale=0.25, angle=0]
                          {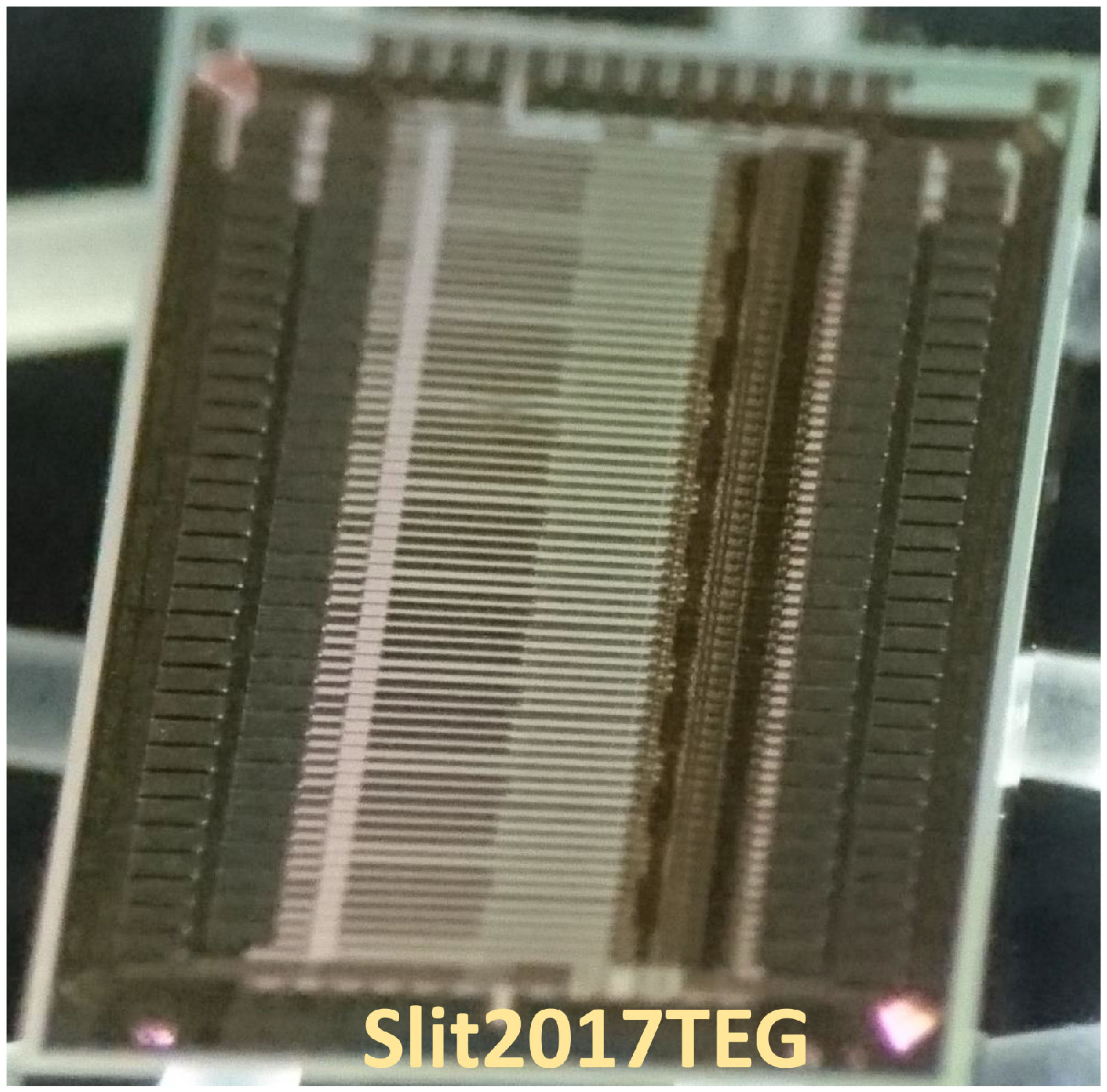}
       \end{minipage}
 
 
      \begin{minipage}{0.02\hsize}
      \end{minipage}

      \begin{minipage}{0.7\hsize}
          \includegraphics[keepaspectratio, scale=0.28, angle=0]
                          {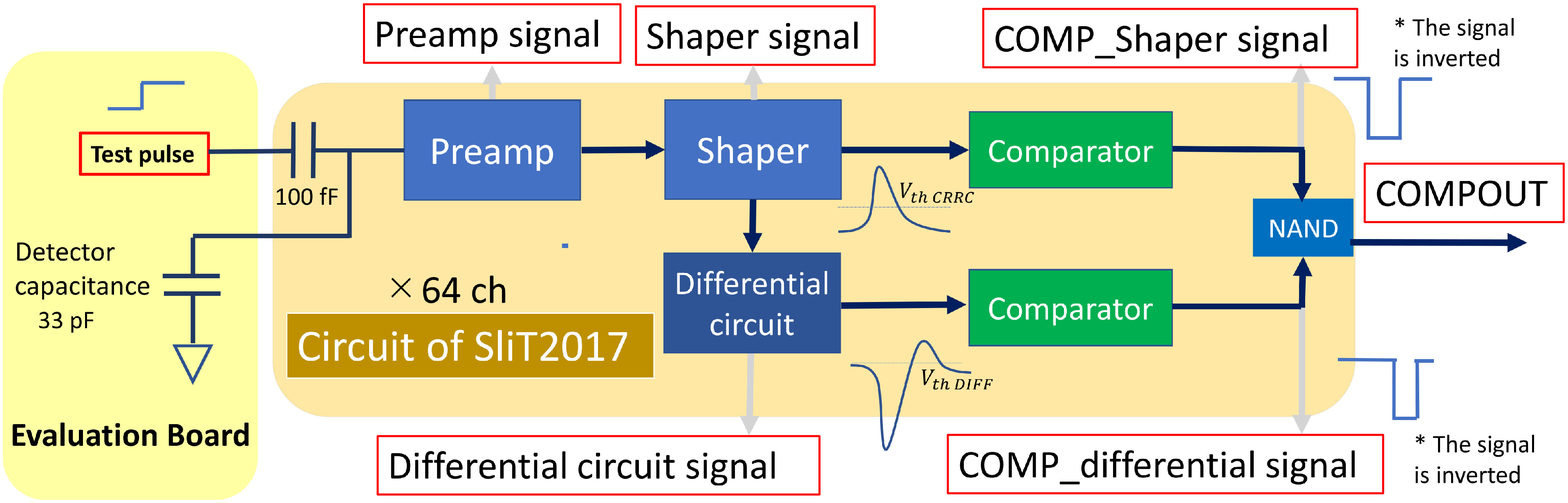}
      \end{minipage} \\
    \end{tabular}
                             \caption{Photograph (left) and the block diagram (right) of the prototype ASIC.}
                          \label{fig_chip}

\end{figure*}

The front-end ASICs for silicon-strip sensors are demanded to tolerate a high hit rate of 1.4~MHz per strip, and at the same time, to have an operational stability under the changing hit-rate by a factor of $1/150$. These experimental requirements can be translated to concrete circuit specifications summarized in Table \ref{tab1}. In a previous study, we have evaluated an analog prototype ASIC and succeeded to operate module-prototype chips ``SliT128A''. This module ASIC includes 128 readout channels, buffer memory, and other digital processing circuits to store and handle timing information \cite{sato, ueno}. The critical performance of the time-walk, however, remained at 12.3~ns and seemed to be difficult for further improvement to reach the requirement of less than 5~ns. Here, we define the time-walk as a timing difference of the rising edges between the 0.5 and 3~MIP events. In order to meet the specification, we newly designed an analog prototype ASIC ``SliT2017TEG'',  implemented with the Silterra 180 nm CMOS technology. Figure \ref{fig_chip} shows a photograph and a block diagram of the ASIC. The chip size is 3880~$\mu$m by 2750~$\mu$m, including 64 readout channels with a 50~$\mu$m channel pitch and staggered bonding pads. 

Each readout channel consists of a capacitor for test pulse injection, a charge-sensitive amplifier (CSA), a CR-RC shaper, a differentiator, two comparators, and control registers for tuning thresholds. A main difference from the previous architecture is in adding a differentiator at the CR-RC shaper. In the previous ASICs, the hit timing was produced from a first-order CR-RC shaper output with a peaking time of 50~ns and a comparator. This method, however, is not able to lower the time-walk of comparator's rising edge below the requirement of 5~ns, even after the fine threshold tuning. In order to improve the performance for an input range of 4~MIPs, i.g., $Q_{\rm sig}=3.84$~fC per MIP, we use a differentiator to shape the semi-gaussian signal into a bipolar-one. The merit to use this architecture is that the cross timing of a bipolar's baseline becomes independent of input charges and it stays almost at the peaking time of the CR-RC output. As a result, the comparator's rising edge from the differentiator's path remains $<50$~ns, and at the same time, the time-walk can be considerably improved. The noise level after the differentiator, i.e., a bipolar shape, becomes worse than the ENC at the CR-RC shaper, we thus preserve the CR-RC shaper path and use both comparators'  information for final pulses by making an NAND operation.

\section{Performance evaluation of ASIC}
\begin{figure*}[!t]
  \centering
    \begin{tabular}{c}

      \begin{minipage}{0.48\hsize}
          \includegraphics[keepaspectratio, scale=0.4, angle=0]
                          {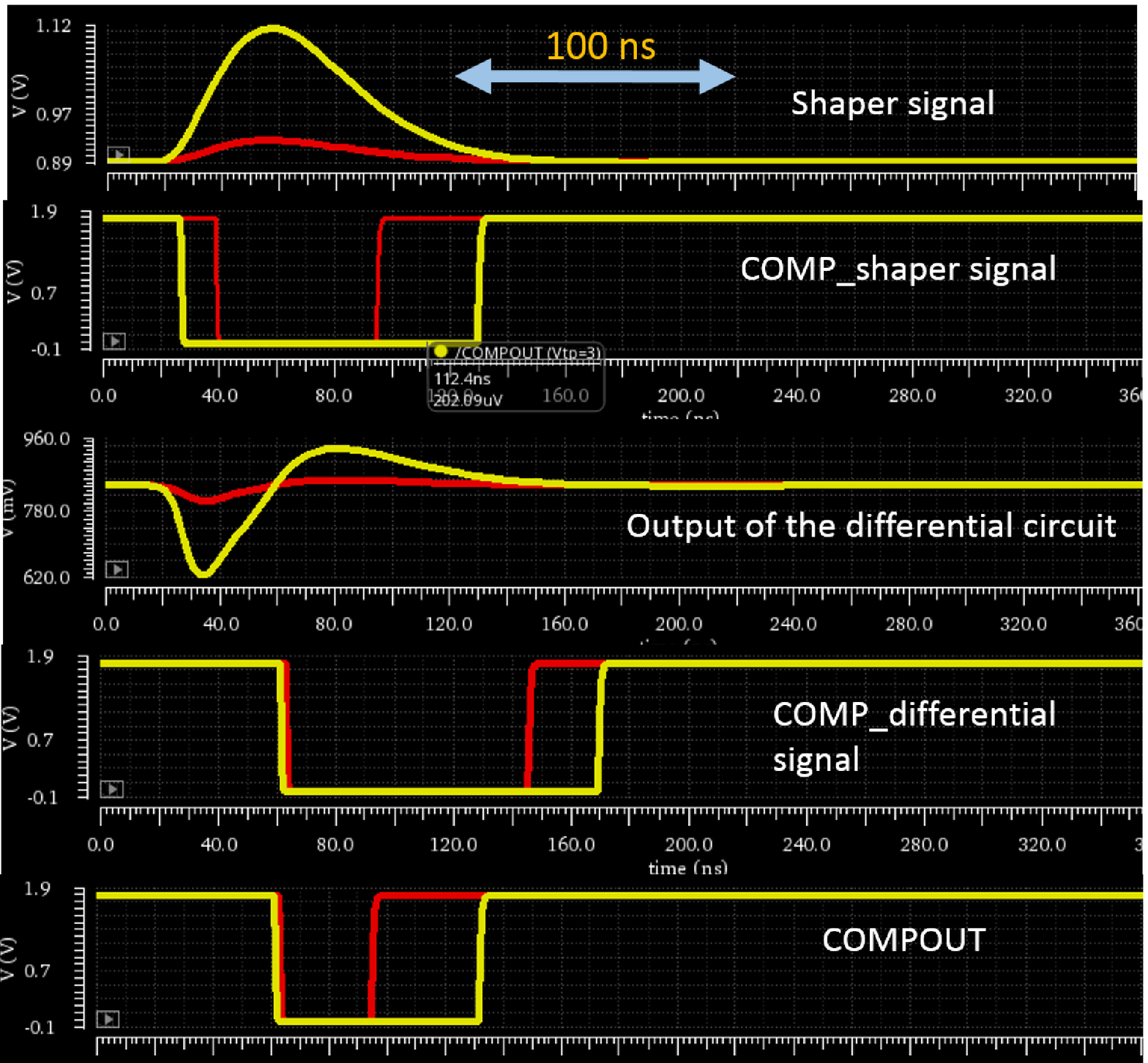}
       \end{minipage}
 
 
      \begin{minipage}{0.02\hsize}
        \hspace{2mm}
      \end{minipage}

      \begin{minipage}{0.48\hsize}
          \includegraphics[keepaspectratio, scale=0.4, angle=0]
                          {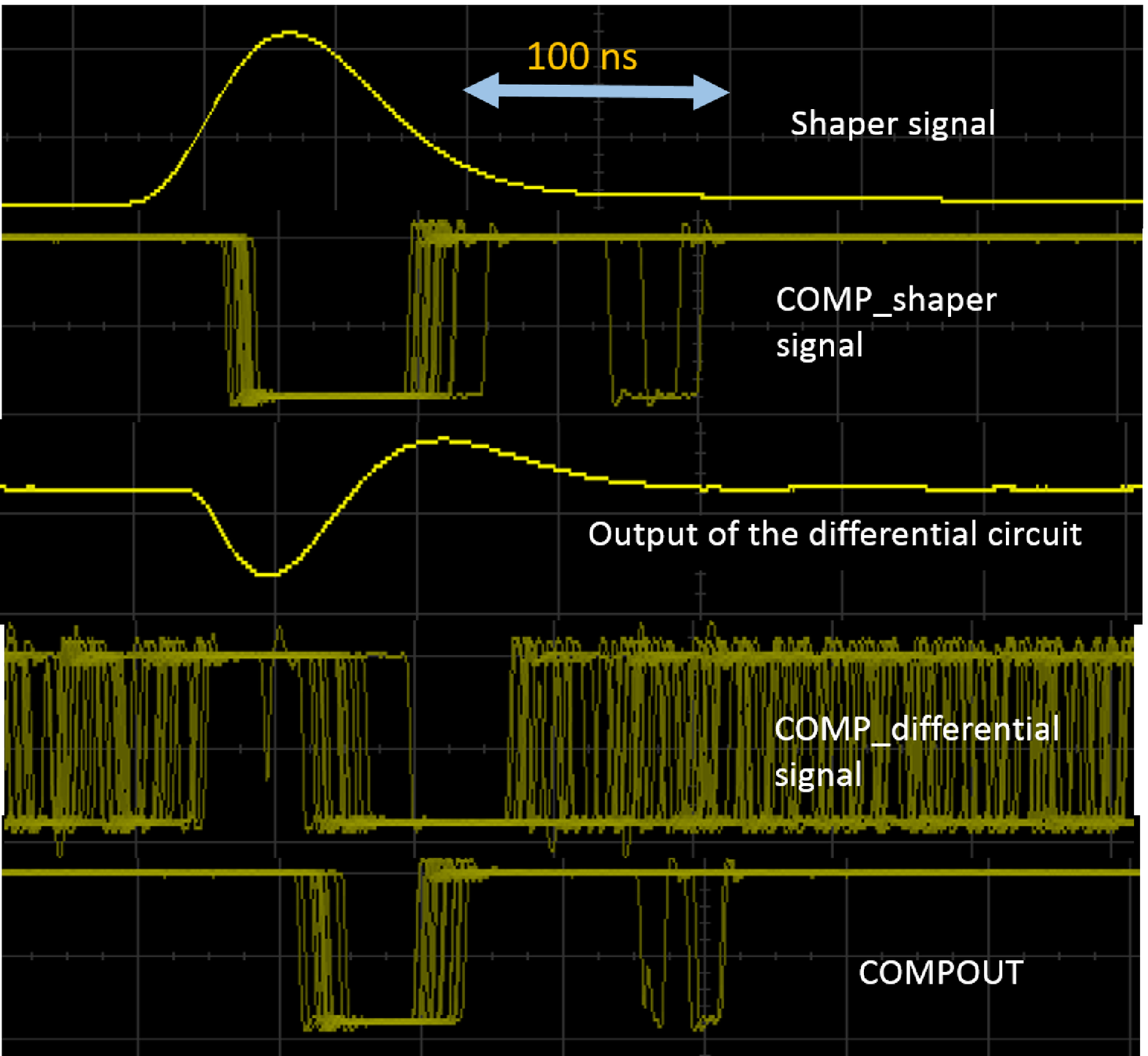}
      \end{minipage} \\
    \end{tabular}
                             \caption{Waveforms from the SPICE simulation (left) and measurement (right). In the left panel, the different injected charges are shown in different colors, i.e., 0.5~MIP (red), and 3~MIP (yellow). The outputs from the CR-RC shaper and the differentiator are averaged for viewability. For the COMPOUT signal the time jitter of the leading edge is much larger than the time walk. The SliT2017TEG does not satisfy the required noise value causing jitter, but the noise of SliT128B manufactured for the actual machine satisfies the S / N ratio of 15 or more. Furthermore, unlike the time walk, jitter can be reduced by taking the average.
}
                          \label{fig_waveform}

\end{figure*}

We evaluated the performance of SliT2017TEG with test pulses generated by a function generator. The ASIC was directly mounted on the PCB and bias currents to the chip were provided by potentiometers.
Figure \ref{fig_waveform} shows the waveform comparison between the SPICE simulation and measurement. From an upper panel, outputs from the CR-RC shaper, the comparator from the CR-RC path, the differentiator, the comparator from the differentiator path, and the NAND gate are shown. 
The measured pulse height and peaking time are consistent with the SPICE simulation.
By optimizing the threshold, i.e., $V_{\rm th, CR-RC}=0.3$~MIP and $V_{\rm th, diff}$ is a few mV above the baseline, we succeeded to achieve the time-walk of 0.8~ns in the measurement for the input range of above 6~MIPs. The measurement results are summarized in Table \ref{tab1}.
The comparator output seems chattering in the right panel. We confirmed that this phenomenon is due by cross-talks between the digital CMOS outputs and analog input line. In order to improve this issue, we will implement a LVDS driver as output buffers. The comparator's outputs are normally handled by a digital circuit and the output buffer is basically implemented for a debugging purpose, the overall performance has no effects in the module production chips.   
The ENC of the measurement remains a bit higher than the requirement. This issue can be improved by adding external decoupling capacitors on the PCB, i.e., 0.1 $\mu$F, to bias nodes of the preamplifier. Although the current ASIC has no access to these bias nodes, we expect a noise performance of 1180~$e^-$ in the SPICE simulation. 
The module production ASIC is currently under preparation. 

\begin{table*}
\begin{center}
\caption{\label{tab1} Requirement and measured performance of the prototype ASIC.}
\small
  \begin{tabular}{|l|l|l|l|} \hline
Parameter & Requirements & Simulation & Measurement  \\ \hline
Peaking time & $<50$ ns & 48.04~ns & 48~ns \\
ENC@$C_{\rm det}=$30~pF & $<1600$~e$^-$ (if S/N>15) & 1661~e$^-$ & 1900~e$^-$@$C_{\rm det}$=33~pF \\
Dynamic range & $>4$~MIP & 8~MIP & $>6$~MIP \\
Pulse width (1 MIP) & $<100$~ns & 48.15~ns & 45~ns \\
Time walk & $<1$~ns & 0.4~ns & 0.8~ns \\ \hline
  \end{tabular}
\end{center}
\end{table*}

%
%
%

\section{Conclusion}
We are developing an ASIC to read the sensor signal for the J-PARC muon g-2/EDM experiment. 
The experiment is aimed to measure the muon anomalous magnetic moment with a precision of 0.1~ppm and to search for electric dipole moment with a sensitivity of 10$^{-21}e\cdot$cm.
As a readout electronics from silicon-strip detectors, we have developed dedicated front-end ASICs.
The prototype analog ASIC was newly submitted to improve the time-walk performance of less than 5~ns. The main difference from the previous architecture is adding a differentiator at the output of a semi-gaussian CR-RC shaper. As a result, the semi-gaussian signal becomes bipolar, and zero-cross timing becomes independent from the input charges. By combining  the timing information produced from a CR-RC shaper output and differentiator output we succeeded to reach the time-walk of 0.8~ns for an input range of 6~MIPs. 
Based on these encouraging results, production chips for the real machine are under preparation. In the final chip, the binary signal from the comparator is sampled with 5~ns intervals for the period of 40.96~$\mu$s and saved in a memory buffer. The data is read out by a serial communication protocol.
The detector module production is scheduled to be completed until May 2019.

\acknowledgments
This work was supported by the JSPS KAKENHI (Grant No. 17H04840).


\end{document}